\documentclass[10pt, a4paper, copyright, logo, nonumbering]{instadeep}

\title{Machine Learning Interatomic Potentials: library for efficient training, model development and simulation of molecular systems}

\correspondingauthor{c.brunken@instadeep.com, j.tilly@instadeep.com}


\author[1]{Christoph Brunken}
\author[1]{Olivier Peltre}
\author[1]{Heloise Chomet}
\author[1]{Lucien Walewski}
\author[1]{Manus McAuliffe}
\author[1]{Valentin Heyraud}
\author[1]{Solal Attias}
\author[1]{Martin Maarand}
\author[1]{Yessine Khanfir}
\author[1]{Edan Toledo}
\author[1]{Fabio Falcioni}
\author[1]{Marie Bluntzer}
\author[1]{\\ Silvia Acosta-Guti\'{e}rrez}
\author[1]{Jules Tilly}

\affil[1]{InstaDeep}

\begin{abstract}
Machine Learning Interatomic Potentials (MLIP) are a novel \textit{in silico} approach for molecular property prediction, creating an alternative to disrupt the accuracy/speed trade-off of empirical force fields and density functional theory (DFT). In this white paper, we present our MLIP library which was created with two core aims: (1) provide to industry experts without machine learning background a user-friendly and computationally efficient set of tools to experiment with MLIP models, (2) provide machine learning developers a framework to develop novel approaches fully integrated with molecular dynamics tools. The library includes in this release three model architectures (MACE, NequIP, and ViSNet), and two molecular dynamics (MD) wrappers (ASE, and JAX-MD), along with a set of pre-trained organics models. The seamless integration with JAX-MD, in particular, facilitates highly efficient MD simulations, bringing MLIP models significantly closer to industrial application.
The library is available on \href{https://github.com/instadeepai/mlip}{GitHub} and on \href{https://pypi.org/project/mlip/}{PyPI} under the Apache license 2.0.
\end{abstract}

\begin{document}
\maketitle
\tableofcontents

% It is good practice to split the content of the document into seperate Tex files,
% and inport them as needed.
% \input{content/01_introduction}

% Example of a figure with its caption

\section{Introduction } \label{sec: introduction}

Evaluation of molecular interactions and properties is critical across multiple sectors, including the pharmaceutical, chemical, and materials industries. Because experimental evaluations are often costly and time-consuming, \textit{in silico} methods have become essential for screening and prioritizing candidate systems. The two main approaches used in research and industry are empirical force fields and quantum chemistry methods. Empirical force fields offer high efficiency, but can fall short in accuracy and fail to capture chemical reactivity. Quantum chemistry methods, while considered the gold standard for accuracy, often remain too computationally intensive for large-scale or routine use.

Machine Learning Interatomic Potential (MLIP) models aim to approach the accuracy of quantum chemistry methods at a fraction of the computational cost. Because they balance speed and accuracy, MLIP models, like traditional force fields and quantum methods, operate within an inherent trade-off between efficiency and precision. This creates space for a wide variety of model architectures, inductive biases, and scales, tailored to different simulation needs. Contributing to this diversity, some models can be fine-tuned for specific systems, while others prioritize broad generalizability across a wide range of chemical space. 

In this white paper, we present a unified framework for MLIP training and deployment in molecular dynamics (MD) simulations. Our objective is to provide a versatile toolkit that serves users across a wide range of backgrounds, from computational chemistry researchers to machine learning developers. The library enables those with minimal machine learning experience to run simulations with just a few lines of code, while offering advanced users the flexibility to develop custom methods with seamless integration into existing training and simulation workflows. We also include a set of models, pre-trained for organic chemistry, that can be readily deployed for simulations or further fine-tuned for specific use cases.

Although many studies have demonstrated that MLIP can achieve near-DFT accuracy while being orders of magnitude faster, the field faces stiff competition from traditional force fields due to their efficiency and scalability \cite{stevens2023molecular,shaw2021anton3}. We therefore see inference speed as a critical area of focus for any MLIP library oriented on usability and a key component for the future success and applicability of these methods. As such, the library is entirely JAX-based \cite{jax2018}, benefiting from full just-in-time XLA (Accelerated Linear Algebra) compilation. In particular, the efficient integration of the MLIP models with the JAX-MD simulation backend allows for state-of-the-art MD simulation speeds.

\section{Brief overview of MLIP methods} \label{sec: lit_review}

\subsection{Related work}

To put this library in the context of the MLIP field, we have briefly referenced below a number of relevant methods. This section is meant as an illustration of the field, rather than aiming to be exhaustive. The idea of using neural networks as force fields stems from the observation that traditional empirically fitted potentials are limited in their functional form and expressivity. As such, many functional forms and inductive biases have been tried over the years, with the field evolving from system specific force fields to more generalized methods covering large portions of the periodic tables in recent years \cite{batatia2024foundationmodelatomisticmaterials, levine2025openmolecules2025omol25}.

\paragraph{Kernel based methods:} Gaussian Approximation Potentials (GAP) \cite{Bartk2010,bartók2020gaussianapproximationpotentialsbrief, Bartk2018} are among the first significant breakthroughs in the field, allowing bespoke high-accuracy training on specific systems while still suffering from a lack of generalizability. From a given set of atomic descriptors, GAP is fitted using a gaussian process regression to simulate interatomic potential energy surfaces. GAP would normally rely on the Smooth Overlap Atomic Positions (SOAP) descriptors \cite{Bartk2013} (other descriptors have been developed and used in GAP (\citet{Behler2007,Thompson2015}), which take into account the geometry of the local atomic environment to compute a set of features for each atomic node used in the kernel function.

The symmetry-adapted Gradient Domain Machine Learning (sGDML) \cite{chmiela2017, chmiela2018, chmiela2020, chmiela2023} is an alternative kernel based method, although it presents significant design differences when compared to GAP, it allows for far greater generalizabilty. sGDML does not use pre-computed descriptors and relies directly on molecular geometry as input. Unlike most MLIP methods, it operates in the gradient domain, meaning it focuses on learning forces rather than energy. Energy conservation is ensured by the construction of the vector-value kernel function used to predict forces. 

\paragraph{Linear methods:} An alternative to kernel methods is to handcraft specific descriptors of a molecular system and use a linear combination of generalized basis functions to predict the potential energy surfaces. This approach has two main advantages over kernel methods: better generalization and, in general, better scaling with respect to system size. Ralf \citet{Drautz2019} showed that many of the atomic centered descriptors (incl. SOAP) are specific instances of a general polynomial expansion of the atomic neighbor density labeled as Atomic Cluster Expansion (ACE). Alternative examples include Moment Tensor Potentials (MTP) \cite{Shapeev2016}.

\paragraph{Atomic Environment Vectors:} The ANI (Accurate NeurAl networK engINe for Molecular Energies) family of models (e.g., ANI-1 \cite{Smith2017}, ANI-1x \cite{Smith2020}, ANI-1ccx \cite{Smith2019}, ANI-2x \cite{Devereux2020}) proposes a different approach, where feed forward neural networks are trained on handcrafted Atomic Environment Vectors (AEVs, adapted from the basis functions proposed by \citet{Behler2007}) to predict molecular energies and forces at near-DFT accuracy (and at times near Coupled Cluster accuracy \cite{Smith2019}. While specialized for small organic molecules, ANI models have been showed to offer good transferability and computational efficiency, and have placed themselves as a reference benchmark for multiple specialized applications at times outperforming DFT (B3LYP) and OPLS \cite{Rezaee2024}. 

\paragraph{Graph Neural Networks:} GNNs or message passing networks form an alternative to bespoke-designed descriptors used in kernel-based methods. In general, GNNs for MLIP are constructed such that the network uses message passing (or at least, local graph neighbors information passing) to learn a representation of the atomic environment, which includes atomic species and geometry of the neighboring atomic structure. These learned descriptors are then usually passed through a simple readout block to predict local energy contributions. A key aspect of graph based networks is that they are constructed to preserve specific system symmetries - at the very least invariance to rotation and translations in energy predictions, but oftentimes also equivariance of spatial output and latent information throughout the network. While equivariance usually comes at a computational cost, it has also been showed to improve data efficiency \cite{Batzner2022, brehmer2024doesequivariancematterscale}. 
We outline below the main categories of graph-based MLIPs:
\begin{itemize}
    \item \textbf{Distance-based (invariant) GNNs:} A first approach is to rely on interatomic distances to learn rotationally invariant energy predictions, such as in SchNet \cite{Schtt2018, schutt2019schnetpack, schutt2023schnetpack}. Similar ideas were developed with a direct focus on periodic crystal structures, such as the Crystal Graph Convolutional Neural Network (CGCNN) \cite{Xie2018}. Despite the more popular equivariant methods outlined below, some invariant methods (such as AIMNet2 \cite{Anstine2025}) have shown excellent accuracy and efficiency on specialized families of systems, while also including charge information in training and predictions\\
    \item \textbf{Directional / angular equivariant GNN:} A first example approach to achieve equivariance in GNN was proposed by \citet{satorras2022enequivariantgraphneural}, and involves equivariantly updating edge features between each message passing layers. Other approaches have alternatively proposed to guarantee inter-layer equivariance through computation of angular features, such as presented in DimeNet \cite{gasteiger_dimenet_2020, gasteiger_dimenetpp_2020}, GemNet \cite{GemNet, gasteiger2024gemnetuniversaldirectionalgraph}, and ViSNet \cite{Wang2024visnet, Wang2024ai2bmd}. \\
    \item \textbf{E(3)-Equivariant GNN / steerable 3D convolutions:} An alternative approach is to build upon the formalism of Clebsch-Gordan-based steerable 3D convolutions \cite{weiler20183dsteerablecnnslearning, kondor2018clebschgordannetsfullyfourier, batatia2022designspacee3equivariantatomcentered} to achieve arbitrary orders of representation of geometric features. Examples of models in this category include NequIP \cite{Batzner2022}, largely based on the Tensor Field Network architecture \cite{thomas2018tensorfieldnetworksrotation}, Allegro \cite{Musaelian2023}, a fully-local (non-message passing) version of NequIP designed for efficient parallelization \cite{musaelian2023scalingleadingaccuracydeep}, and MACE \cite{batatia2023macehigherorderequivariant, kovács2025maceofftransferableshortrange, batatia2024foundationmodelatomisticmaterials, Kovcs2023}, which formalizes the connection between the steerable convolution and the ACE basis by constructing a learnable multi-body atomic cluster expansion. PaiNN (Polarizable Atom Interaction Neural Network)\cite{schütt2021equivariantmessagepassingprediction} offers a more efficient alternative relying solely on vector features rather than full tensor algebra. It is worth noting that this category of equivariant GNN largely relies on specialized libraries managing the steerable convolution features \cite{e3nn_lib, unke2024e3x}. Other backends also include accelerated CUDA kernels, such as NVIDIA's \textit{cuEquivariance} package. Finally, some approaches use projections onto 2D domain to perform faster convolutions \cite{passaro2023escn, luo2024enablingefficientequivariantoperations, fu2025learningsmoothexpressiveinteratomic}.

\end{itemize}

\subsection{Models included in the \textit{mlip} library}

In the current version of the library, we have incorporated three graph-based MLIP models: MACE \cite{batatia2023macehigherorderequivariant}, NequIP \cite{Batzner2022}, and ViSNet \cite{Wang2024visnet}. All three were chosen based on their strong performance and extensive validation by the community across diverse settings. 

In the first version of the library, we have maintained model architectures as closely as possible to their original implementation. We aim for later versions to include modified layers and backends (see our roadmap below). However, we will endeavor to maintain backwards compatibility. The code sources and modifications are outlined below: 

\begin{itemize}
    \item \textbf{MACE:} The code for MACE is in large parts based on the \href{https://github.com/ACEsuit/mace-jax}{initial JAX implementation} by Mario Geiger and Ilyes Batatia. We implemented a number of minor changes to match the inference output of \href{https://github.com/ACEsuit/mace}{the original Torch version} as closely as possible, with identical weights. It is worth noting that perfect matching is challenging due to a different activation normalization between \href{https://e3nn.org/}{e3nn} and \href{https://github.com/e3nn/e3nn-jax/}{e3nn-jax}. Additionally, models will differ when used in float32 precision due to different rounding conventions between JAX and Torch. Other changes include additional Flax versions of some modules and minor refactoring.\\
    \item \textbf{ViSNet:} The code for ViSNet was entirely converted to JAX from the \href{https://github.com/microsoft/AI2BMD/tree/ViSNet}{original Torch implementation} and was likewise set to match the inference output for a given set of weights. \\
    \item \textbf{NequIP:} The code for NequIP is almost entirely based on the version implemented in \href{https://github.com/google-deepmind/materials_discovery/blob/main/model/nequip.py}{the GNoME repository} \cite{merchant2023scaling}. Only minor modifications where made to fit within the library workflow. We did not attempt to match NequIP to the \href{https://github.com/mir-group/nequip}{original Torch code}, a slightly modified version of which we used for benchmarking (see the relevant section below).
\end{itemize}

The library is designed to support the seamless addition of new models. To facilitate this, as part of the documentation, we provide a \href{https://instadeepai.github.io/mlip/user_guide/index.html#jupyter-notebook-tutorials}{tutorial} on how to write new models to be interfaced easily with the other parts of the library (e.g. training or simulation). 
Looking ahead, as outlined in our roadmap, we plan to incorporate additional JAX implementations of MLIP models in new releases. 

\section{Library overview} \label{sec:library_overview}

\subsection{Purpose and design philosophy}

The purpose of the \textit{mlip} library is to provide users with a toolbox to deal with MLIP models in a true end-to-end fashion. This includes data preprocessing, implementation of multiple model architectures, model training, model fine-tuning, deployment through MD simulation, energy minimization, and batched inference. The \textit{mlip} library was built in accordance with the following key design principles: 

\begin{itemize}
    \item \textbf{Ease-of-use:} The library should be simple to install and use, especially for non-expert users, who primarily aim to apply pre-trained MLIP models to relevant scientific applications and may have limited prior experience with the JAX ecosystem. Furthermore, we are aware that ML models and workflows typically rely on a large number of configurable values. Hence, we provide sensible default parameters wherever possible without unnecessarily reducing flexibility for users who can take advantage of it. \\
    \item \textbf{Extensibility:} For more experienced users, we want \textit{mlip} to be a toolbox that can be extended easily. For example, users can seamlessly complement the library by adding a new model architecture, alternative data preprocessing methods, or additional simulation backends. Hence, we embrace the modularity of these components in the library design wherever possible.\\
    \item \textbf{Inference efficiency:} We believe that to successfully push MLIP models towards relevant industrial applications, high inference speeds are essential. Most relevant applications, especially in biology, rely on running long MD simulations on large systems. Therefore, we aim to deliver the most efficient model implementations and simulation pipelines. We prioritize inference over training speed when necessary. However, we strive to be state-of-the-art in both areas.
\end{itemize}

\begin{figure}[ht]
  \centering
  % \fbox{\rule[-.5cm]{0cm}{4cm} \rule[-.5cm]{4cm}{0cm}}
  \includegraphics[width=\linewidth]{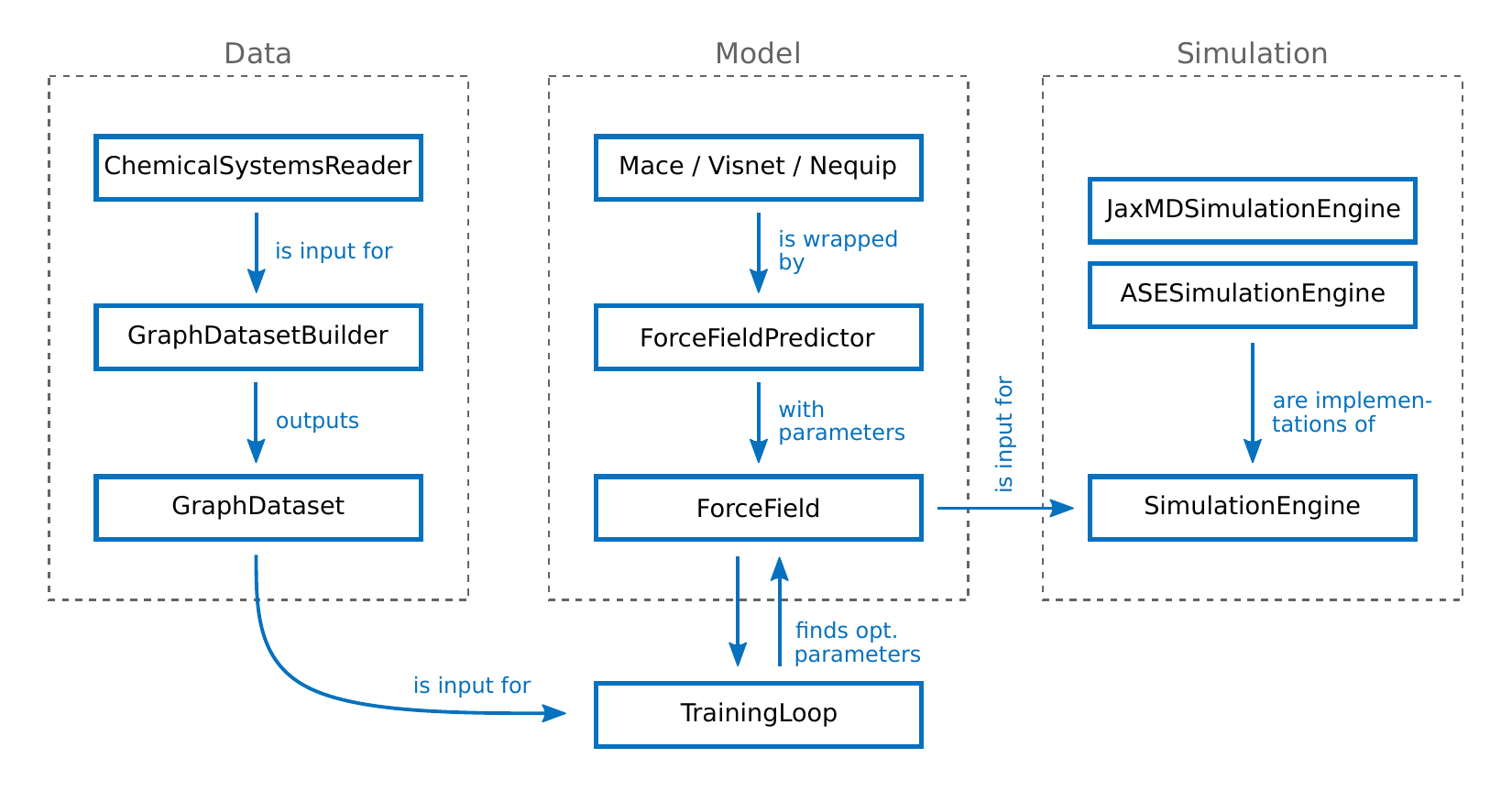}
  \caption{Schematic overview of the essential classes of the \textit{mlip} library and their interactions. To run simulations with one of the two implementations of \texttt{SimulationEngine}, we need to input an instance of \texttt{ForceField}, which contains a \texttt{ForceFieldPredictor} (implemented as a \textit{Flax} \cite{flax2020github} module) and its parameters. A \texttt{ForceField} instance can be called directly on an input graph. To train the force field model, i.e., find its optimal parameters, we provide a \texttt{TrainingLoop} class that requires training and validation data in \texttt{GraphDataset} objects. These objects can be created easily with tools in the \texttt{data} module. See the \href{https://instadeepai.github.io/mlip/user_guide/index.html}{tutorials} in the code documentation for more details.}
  \label{fig:library_overview}
\end{figure}

\subsection{Structure and modules}

The \textit{mlip} library is constructed in a modular way, separating model implementation from training, fine-tuning, and simulation code. It consists of multiple sub-modules targeted towards different parts of a full MLIP pipeline.

\begin{enumerate}
\item The \textbf{\texttt{data}} module contains code related to dataset preprocessing. Its main purpose is to go from datasets stored on a file system to instances of \texttt{GraphDataset} classes that can be directly used for training or batched inference tasks.

\item The \textbf{\texttt{models}} module contains code related to the MLIP models, i.e., their core implementations, loss classes, and other related utilities, such as loading of trained models.

\item The \textbf{\texttt{training}} module contains code related to training or fine-tuning MLIP models.

\item The \textbf{\texttt{simulation}} module contains code for running MD simulations or energy minimization with MLIP models. We support both JAX-MD \cite{schoenholz2020jaxmdframeworkdifferentiable} and ASE \cite{HjorthLarsen2017} backends.

\item The \textbf{\texttt{inference}} module contains a function to run batched inference on a list of structures with MLIP models.

\item The \textbf{\texttt{utils}} and \textbf{\texttt{typing}} modules contain utility functions, data classes, and type aliases used in other modules that may also be useful for various downstream tasks.
\end{enumerate}

Each of these modules is designed to allow the user to set up their own experiment scripts or notebooks with minimal effort, while also supporting customization, especially for topics such as logging (e.g. to a remote storage location like Amazon S3 or Google Cloud Storage) or adding new losses, MLIP model architectures, or dataset readers.

In Figure~\ref{fig:library_overview}, we provide a schematic overview of the most essential classes of the library and how they interact with each other.

\subsection{Practical examples}

The \textit{mlip} package can be installed via \texttt{pip}. We provide full code \href{https://instadeepai.github.io/mlip}{documentation} with many tutorials on how to use the library. 
 In the following, we present two common use cases: (1) launching an MD simulation with one of the pre-trained models, and (2) training a model from scratch. These two examples aim to provide a general overview of the library API. For a complete step-by-step walkthrough, please refer to the \href{https://instadeepai.github.io/mlip/user_guide/index.html#deep-dive-tutorials}{tutorials}.

In the first example, we load a pre-trained MACE model from a zip archive. It is directly loaded into a \texttt{ForceField} object containing all the relevant information about the model. In a subsequent step, we load a chemical system with ASE, initialize the MD config and engine objects, and then launch the run. For more details on logging and results collection, see the \href{https://instadeepai.github.io/mlip/user_guide/simulations.html}{deep-dive simulation tutorial} provided in the code documentation.

\begin{tabular}{p{0.52\textwidth} p{0.45\textwidth}}
\begin{lstlisting}
import ase.io
from mlip.models import Mace, ForceField
from mlip.models.model_io import (
    load_model_from_zip
)
from mlip.simulation.jax_md import (
    JaxMDSimulationEngine
)

# Load pre-trained model
force_field = load_model_from_zip(
    Mace, "/path/to/pretrained_model.zip"
)

# Set up MD prerequisites
atoms = ase.io.read("/path/to/xyz/or/pdb/file")
md_config = JaxMDSimulationEngine.Config(
    num_steps=1_000_000,
    # ... other settings
)

# Run MD
md_engine = JaxMDSimulationEngine(
    atoms, force_field, md_config
)
md_engine.run()

\end{lstlisting}
&
\begin{itemize}
  \setlength\itemsep{.5em}
  \item \texttt{imports}: Load the necessary modules.
  \item \texttt{load\_model\_from\_zip}: Loads a pre-trained MACE model from a zip archive into a \texttt{ForceField} object containing all relevant information about the model. Note that the \texttt{ForceField} class can also be viewed as a generic interface for any JAX function that maps \texttt{jraph.GraphsTuple} graphs to our \texttt{Prediction} objects.
  \item \texttt{ase.io.read}: Reads the structure file (as XYZ, PDB, etc.) using ASE.
  \item \texttt{JaxMDSimulationEngine.Config}: Configures the simulation (e.g., number of steps).
  \item \texttt{JaxMDSimulationEngine}: Sets up the molecular dynamics engine.
  \item \texttt{md\_engine.run}: Launches the MD simulation.
\end{itemize}
\\
\end{tabular}

 Although we also provide ASE as a simulation backend, we recommend to rely on the integration with JAX-MD for simulations wherever possible (as in the example above), as it enables running MLIP-based MD with state-of-the-art speed (see benchmarking results of pre-trained models below). With JAX-MD, we can run a collection of multiple MD steps in a fully JIT-compiled manner on the GPU without any data transfer required between CPU and GPU. During this time, any form of logging or saving of intermediate results is not possible. As a consequence, we separate the JAX-MD based simulations into multiple episodes, where logging happens only between two of them. 
 
 Furthermore, note that JAX has to recompile the force field prediction function each time its input shapes change, for example, caused by a change in the number of edges resulting from a change in atomic positions. To limit the number of times that JAX must recompile, we apply padding to the neighbor lists and check whether the amount of padding is still sufficient after each episode. If the edge buffer overflowed, we reallocate the neighbor lists and rerun the previous episode. Note that the alternative \texttt{ASESimulationEngine} has an analogous interface to the \texttt{JaxMDSimulationEngine}. With ASE, we also use the same padding strategy to avoid recompiling often. However, reallocation is not limited to happening after episodes but can happen after each MD step if necessary. Therefore, in contrast to the \texttt{JaxMDSimulationEngine}, the \texttt{ASESimulationEngine} will not require a number of episodes to be set in its configuration.

The second example is a code snippet to train a model. We first set up all the prerequisites, which include (1) the dataset (see dedicated \href{https://instadeepai.github.io/mlip/user_guide/data_processing.html}{data processing tutorial} in the documentation for more details), (2) the force field model, (3) the loss, (4) the optimizer, and (5) the training loop config. Once these objects have been set up, one can easily instantiate the training loop class and start the training run. Note that we support multi-GPU training via data parallelism. For more details, see the \href{https://instadeepai.github.io/mlip/user_guide/training.html}{model training tutorial} in the code documentation.

\begin{tabular}{p{0.52\textwidth} p{0.45\textwidth}}
\begin{lstlisting}
from mlip.training import (
    TrainingLoop, get_default_mlip_optimizer
)
from mlip.models.loss import MSELoss
from mlip.models import Mace, ForceField

# Get data
train_set, validation_set, dataset_info = (
    _get_dataset()
)

# Initialize model
mace = Mace(Mace.Config(), dataset_info)
force_field = (
    ForceField.from_mlip_network(mace)
)

# Other prerequisites
loss = MSELoss()
optimizer = get_default_mlip_optimizer()
config = TrainingLoop.Config(
    num_epochs=100,
    ...  # other settings
)

# Create TrainingLoop class
training_loop = TrainingLoop(
    train_dataset=train_set,
    validation_dataset=validation_set,
    force_field=force_field,
    loss=loss,
    optimizer=optimizer,
    config=config,
)

# Start the model training
training_loop.run()

\end{lstlisting}
&
\begin{itemize}
  \setlength\itemsep{.5em}
  \item \texttt{imports}: Load the necessary modules.
  \item \texttt{\_get\_dataset}: Placeholder function to be replaced by code to load a dataset from the filesystem. See the \href{https://instadeepai.github.io/mlip/user_guide/data_processing.html}{tutorial} on data reading processing for more information. Model hyperparameters directly related to the dataset are stored in a \texttt{DatasetInfo} object.
  \item \texttt{Mace}: Initialize a MACE model from a configuration and a \texttt{DatasetInfo} object. In this example, the default hyperparameters are used.
  \item \texttt{ForceField.from\_mlip\_network}: Creates the \texttt{ForceField} wrapper class (main interface with training and simulation pipelines) from the MACE network class.
  \item \texttt{MSELoss}: Sets up a loss, in this example, a weighted mean-squared error loss for energy, forces, and stress. In this example, the default weights are used.
  \item \texttt{get\_default\_mlip\_optimizer}: Sets up the default optimizer for MLIP models (see the \href{https://instadeepai.github.io/mlip/api_reference/training/optimizer.html#mlip.training.optimizer.get_default_mlip_optimizer}{code documentation} for more details on how it is implemented).
  \item \texttt{TrainingLoop.Config}: Configures the training loop.
  \item \texttt{TrainingLoop}: Sets up the model training.
  \item \texttt{training\_loop.run}: Launches the model training.
\end{itemize}
\\
\end{tabular}

\newpage
\section{Dataset and pre-trained models} \label{sec: datasets_foundation_models}

In this section, we illustrate the use of the library for large-scale training of MLIP models. To that end, we present a set of three pre-trained models, one for each of the architectures included in the library, which were selected from many training runs. Our aim is to illustrate the usage of the library rather than provide usable models. However, these are nonetheless available under a separate license on \href{https://huggingface.co/collections/InstaDeepAI/ml-interatomic-potentials-68134208c01a954ede6dae42}{InstaDeep's HuggingFace collection}. We describe below the dataset used, the training processes, validation results, and runtime benchmarks.

\subsection{Curated SPICE2 Dataset}

We currently provide access to three pre-trained models, all trained on a curated second version of the SPICE2 dataset \cite{Eastman2023, Eastman2024}. The SPICE2 dataset was chosen for its diversity, both in chemical and conformational space, comprising approximately two million structures computed with DFT at the \(\omega\)B97M-D3(BJ)/def2-TZVPPD level of approximation. The dataset is labeled and subdivided into a collection of subsets; details of the training and validation sets per subset are provided in Table~\ref{tbl:dataset_overview}.

\begin{table}[ht]
  \centering
    \caption{Summary of training and validation sets. Data categories in columns correspond to those in SPICE2 \cite{Eastman2024}.}
  \begin{tabular}{lrrrr}
    \toprule
         & PubChem & DES370K & Amino acid ligand  & Dipeptides \\
    \midrule
    Training set &  1,284,419 & 262,820 & 140,128 & 19,699 \\
    Validation set & 58,336 & 10,917 & 7,250 & 1,250 \\
    Average size & 36.9 & 16.6 & 55.4  & 44.4    \\
    \toprule
              & Monomers & Solvated PubChem &  Water & Solvated amino acid  \\
    \midrule
    Training set & 16,750 &  12,130 & 1,000 & 950 \\
    Validation set & 1,000  & 705 & 0 & 50 \\
    Average size & 13.5 & 92.8 & 95 & 55.4   \\
    \bottomrule
  \end{tabular}
  \label{tbl:dataset_overview}
\end{table}

To improve data quality, several pre-processing steps were applied. First, any structure where a hydrogen atom did not have exactly one chemical bond (according to a detection mechanism based on covalent radii with appropriate tolerance: 0.4 {\AA} on top of the sum of covalent radii) was removed, eliminating 42,689 structures, as these are likely to be unphysical. Next, since these pre-trained models are not designed to handle charged systems, an additional 142,647 non-neutral structures (total charge) were excluded, stabilizing training. Finally, inspired by the filtering strategy used in MACE-OFF \cite{kovács2025maceofftransferableshortrange}, we applied a force filter to remove structures with either a non-zero total force or unusually high per-atom forces. Specifically, we excluded structures with a total force norm exceeding 0.1\,eV/\AA~or any individual force greater than 15\,eV/\AA. Although this filtering represents a tradeoff, improving force prediction accuracy while slightly reducing energy prediction accuracy on the validation set, it led to improved performance on key benchmarks and was therefore adopted. This step removed an additional 1,024 structures.

\subsection{Model training methodology}

For the model training, this dataset was then split into training and validation sets using a 95:5 ratio. The split was performed at the molecular SMILES level, ensuring that different conformers of the same molecule were not included in both sets. As a result, some elements which are rare in this curated version of SPICE2 appear only in the training set but not in the validation set (specifically $\mathrm{K}$, $\mathrm{Li}$ and $\mathrm{Na}$). Although this limitation will be addressed in future updates to these MLIP models, users are currently advised to use caution when applying the models to systems containing these elements. The final training set contains 1,737,896 structures covering 15 chemical elements ($\mathrm{B}$, $\mathrm{Br}$, $\mathrm{C}$, $\mathrm{Cl}$, $\mathrm{F}$, $\mathrm{H}$, $\mathrm{I}$, $\mathrm{K}$, $\mathrm{Li}$, $\mathrm{N}$, $\mathrm{Na}$, $\mathrm{O}$, $\mathrm{P}$, $\mathrm{S}$, $\mathrm{Si}$), while the validation set contains 87,922 structures across 12 elements.

Each pre-trained model was trained for 220 epochs using NVIDIA H100 GPUs. The Visnet and NequIP models were trained using the Huber loss \cite{huber1992}, while the MACE model used the MSE loss. We have detailed below the key parameters, though full details of the model architectures and training hyperparameters can be found in Appendix A.

\begin{itemize}
    \item \textbf{MACE} \cite{batatia2023macehigherorderequivariant, kovács2025maceofftransferableshortrange, batatia2024foundationmodelatomisticmaterials, Kovcs2023}:  The MACE pre-trained model hyperparameters were chosen to prioritize stability of MD simulations. The model has 2 layers and 128 channels. The many-body correlation order $\mathrm{correlation} = 2$ and the degree of node features $\mathrm{node~symmetry} = 3$ (called $\mathrm{max~L}$ in \cite{kovács2025maceofftransferableshortrange}).  
    %\item \textbf{MACE} \cite{batatia2023macehigherorderequivariant, kovács2025maceofftransferableshortrange, batatia2024foundationmodelatomisticmaterials, Kovcs2023}:  The MACE pre-trained model hyperparameters were chosen to align with the MACE-OFF medium \cite{kovács2025maceofftransferableshortrange} model. The model has 2 layers and 128 channels. The many-body correlation order $\mathrm{correlation} = 3$ and the degree of node features $\mathrm{node~symmetry} = 1$ (called max $\mathrm{L}$ in \cite{kovács2025maceofftransferableshortrange}).
    \item \textbf{ViSNet} \cite{Wang2024visnet, Wang2024ai2bmd}: The ViSNet pre-trained model has 4 hidden layers and 128 embedding channels. 8 attention heads were used, as well as 32 RBF features and $L_{\mathrm{max}} = 2$.
    \item \textbf{NequIP} \cite{Batzner2022}: The NequIP pre-trained model uses 5 interaction blocks and $L_{\mathrm{max}} = 2$. The feature configuration is \(64\mathrm{x}0\mathrm{e} + 64\mathrm{x}0\mathrm{o} + 32\mathrm{x}1\mathrm{e} + 32\mathrm{x}1\mathrm{o} + 4\mathrm{x}2\mathrm{e} + 4\mathrm{x}2\mathrm{o}\).
\end{itemize}

These selected settings on MACE notably differ from those used for MACE-OFF medium \cite{kovács2025maceofftransferableshortrange}. This is because we found that, when training on SPICE2, our updated hyperparameters resulted in significantly better MD stability. To provide comparable models trained on the same dataset, we decided to include this version instead of the one aligned to MACE-OFF, despite the additional computational cost and higher energy prediction errors. We present below validation metrics for both MACE (large - our hyperparameters) and MACE (medium - following the MACE-OFF \cite{kovács2025maceofftransferableshortrange} hyperparameters) on SPICE2. We also conducted a training of MACE aligned to the hyperparameters of MACE-OFF on a curated version of SPICE1 \cite{moore_spice1}. We found excellent MD stability and validation metrics (more details are presented in Appendix B). 

Finally, we also trained a model with a modified MACE architecture optimized for inference speed, which we present in detail in Appendix C.

\subsection{Model benchmarking}

\paragraph{Validation results:} We evaluate the performance of the three pre-trained models, MACE-large, NequIP and VisNet, as well as the MACE-medium model (with hyperparameters aligned to MACE-OFF medium in \cite{kovács2025maceofftransferableshortrange}) and the modified MACE model (see Appendix C), on the SPICE2 validation set used during training. Each model was assessed with two standard error metrics: the mean absolute error (MAE) in predicted energies per atom (meV/atom) and the MAE in atomic forces (meV/Å). Validation was conducted across seven subsets of SPICE2, including isolated and solvated small molecules, as described in Table \ref{tbl:dataset_overview}.

\begin{figure}[ht]
  \centering
  % \fbox{\rule[-.5cm]{0cm}{4cm} \rule[-.5cm]{4cm}{0cm}}
  \includegraphics[width=0.95\linewidth]{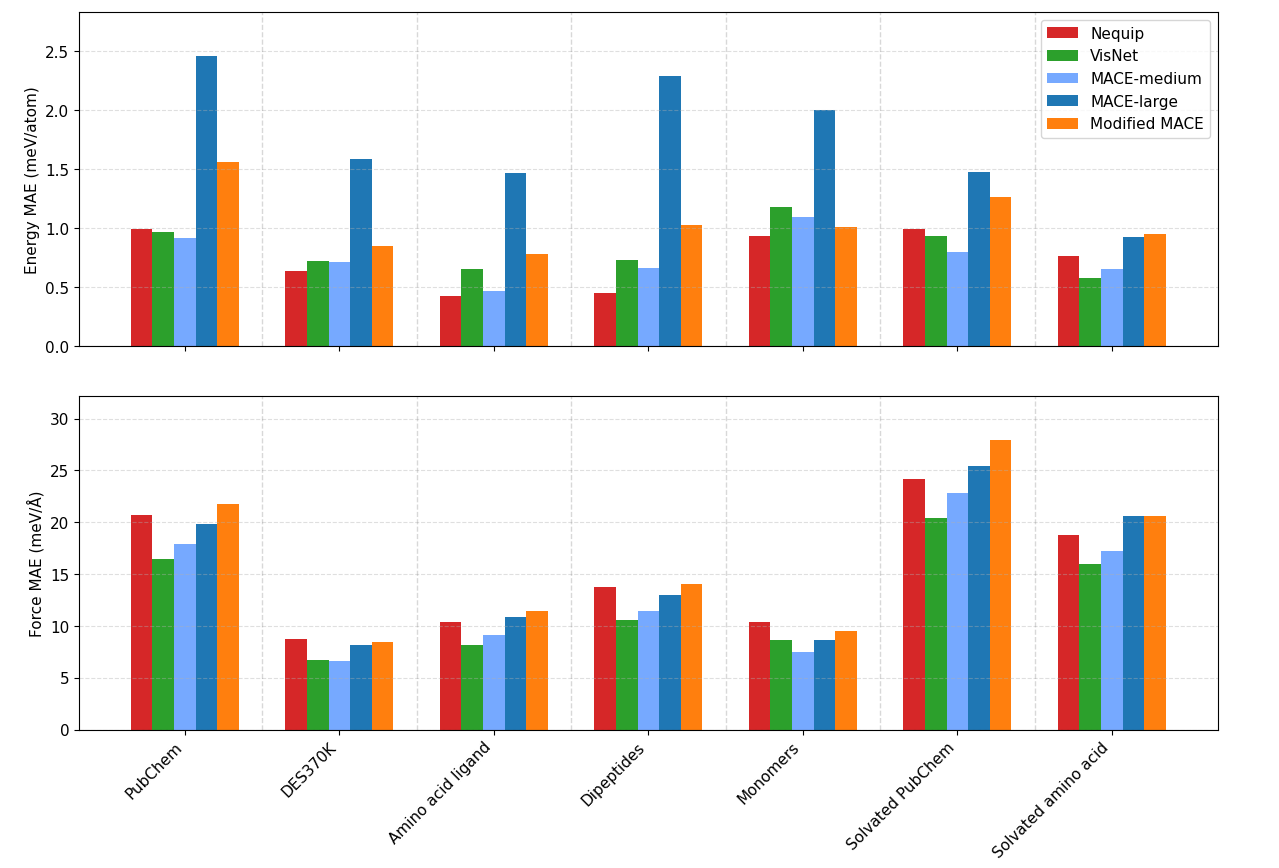}
  \caption{Validation set mean absolute errors (MAE) for energy per atom (meV/atom) and atomic forces (meV/Å) across seven molecular subsets in the SPICE2 dataset. The three pretrained models— MACE-large, VisNet, and NequIP— as well as MACE-medium (following the MACE-OFF \cite{kovács2025maceofftransferableshortrange} hyperparameters) and our modified MACE model (see Appendix C) are evaluated. The subsets include: PubChem, DES370K, amino acid ligands, dipeptides, monomers, solvated PubChem, and solvated amino acids. MAE values reflect the deviation from DFT reference calculations. The number of structures per subset is detailed in Table \ref{tbl:dataset_overview}.}
  \label{fig:validation_metrics}
\end{figure}

Figure \ref{fig:validation_metrics} presents a comparative summary of model performance across all subsets. NequIP achieves the lowest energy MAE for all subsets, while Visnet outperforms in force MAE. Across most models, lower force errors are observed in the DES370K, dipeptides, monomers and solvated amino acids subsets than in the PubChem and solvated PubChem subsets. While MACE-medium achieves lower energy RMSE than MACE-large in every subset, MACE-large, on average, outperforms MACE-medium in force RMSE. Users should be warned that while validation errors are relevant metrics to measure training performance, they are not sufficient to attest to a model's ability to simulate correct physics. 

\paragraph{Runtime benchmark:} Conducting a reliable runtime benchmark can be quite challenging. A first obvious reason is the notable difference in implementation across JAX and Torch versions. As a result, we want to point out that the model implementations on which the Torch + ASE benchmarks are run are our own, and they \textbf{\textit{should not be considered representative of the performance of the code developed by the original authors}}.

\begin{figure}[ht]
    \centering
    \begin{subfigure}[t]{0.38\textwidth}
        \centering
        \includegraphics[width=\linewidth]{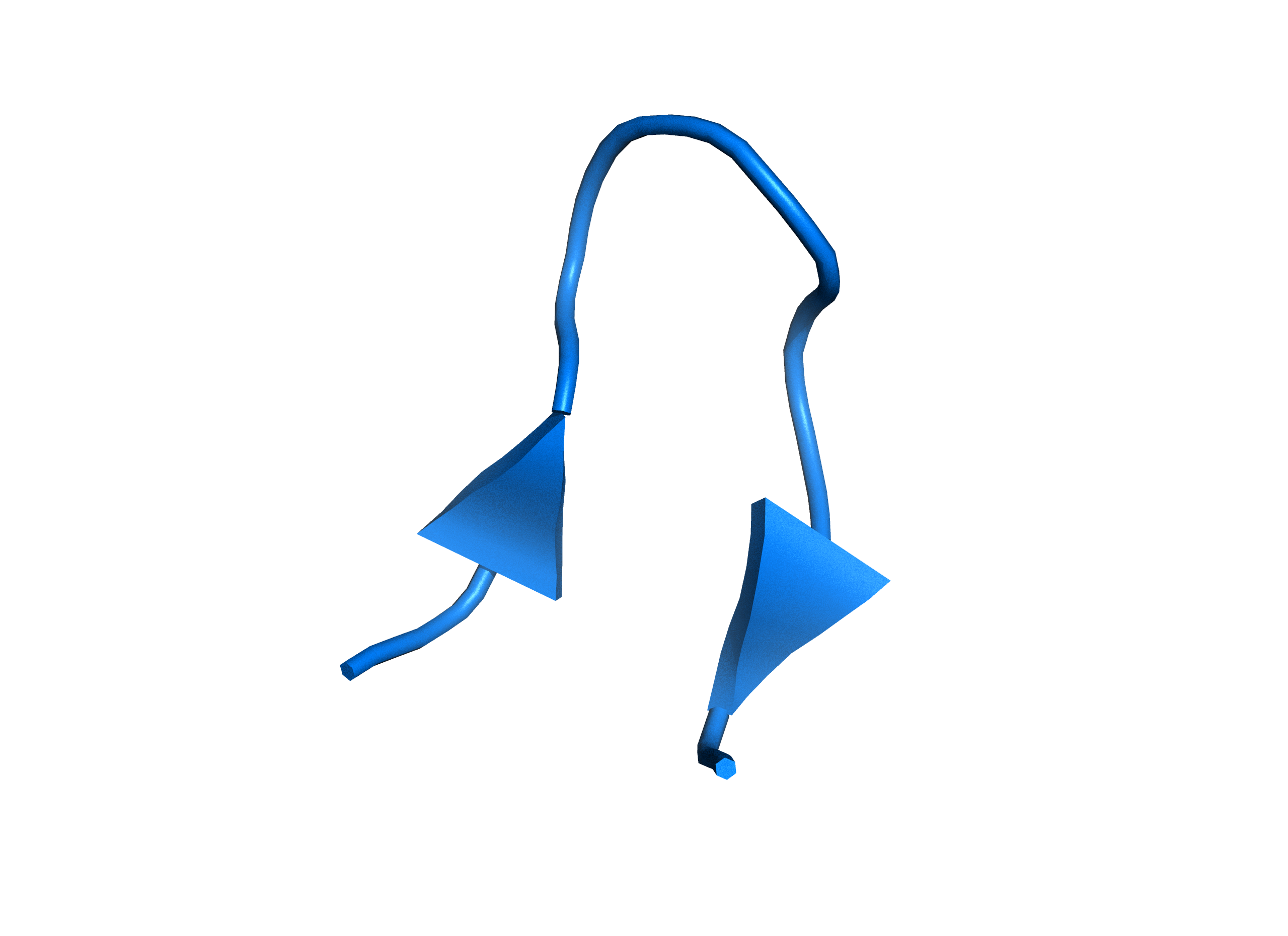}
        \caption{1UAO}
    \end{subfigure}
    \begin{subfigure}[t]{0.48\textwidth}
        \centering
        \includegraphics[width=\linewidth]{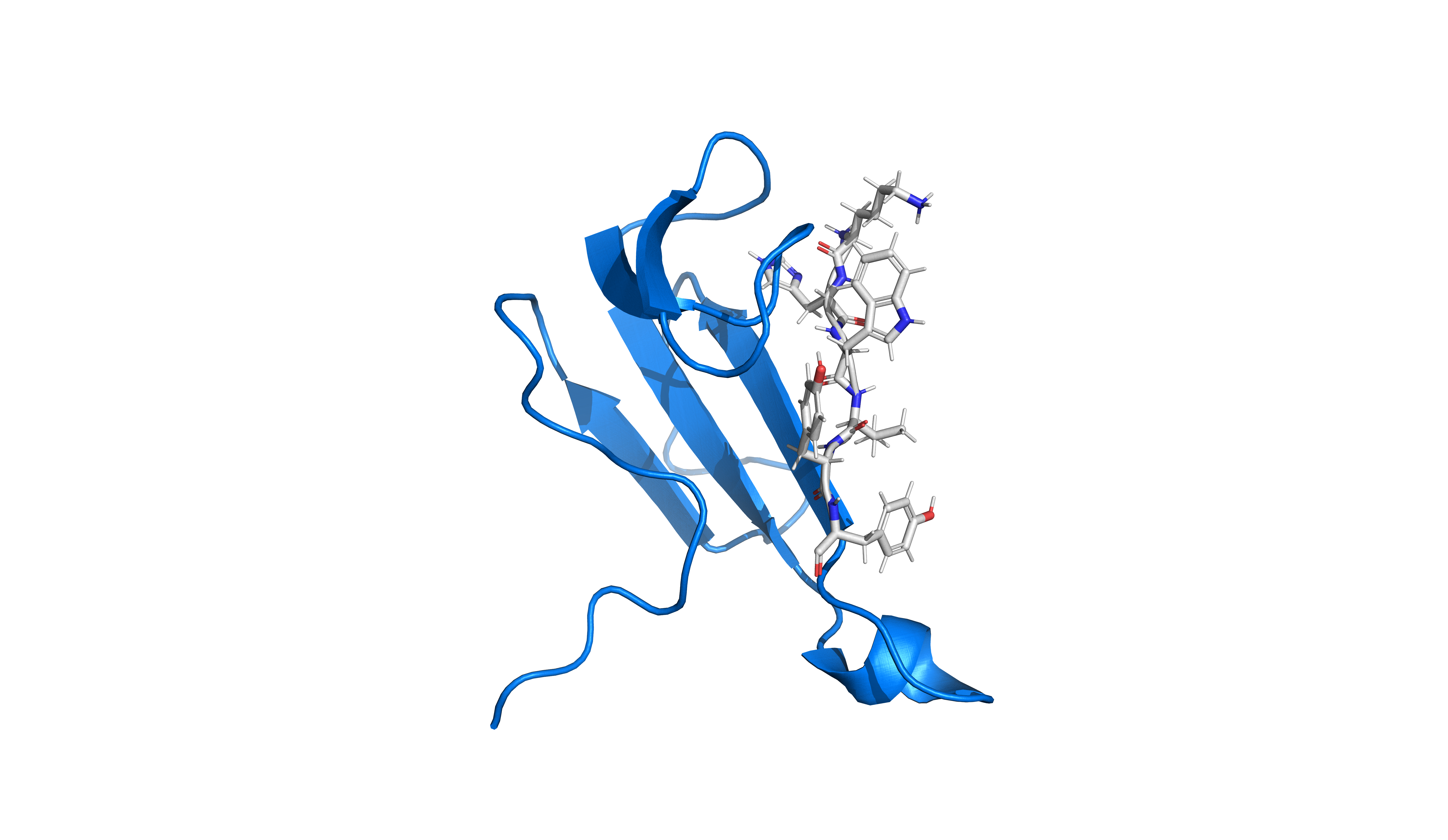}
        \caption{1ABT}
    \end{subfigure}
    \caption{Cartoon representations of benchmark systems: (a) chignolin (PDBid: 1UAO) and the alpha-bungarotoxin in complex with a functionally important region of the nicotin acetylcholine receptor (highlighted in grey)}
    \label{fig:benchmark_structures}
\end{figure}

With this in mind, we present in Table \ref{tbl:models_overview} simulation benchmarks on two different systems (1UAO and 1ABT, see Figure \ref{fig:benchmark_structures}) with results averaged over a 1 nanosecond simulation: 
\begin{itemize}
    \item \textbf{1UAO}: Chignolin (PDBid: \href{https://www.rcsb.org/structure/1UAO}{1UAO}) is a synthetic mini-protein, designed to mimic the $\beta$-hairpin secondary structure motif found in natural proteins. Due to its size, Chignolin is widely used in classical molecular dynamics simulations to study folding. \\

    \item \textbf{1ABT}: Alpha-bungarotoxin complex (PDB: \href{https://www.rcsb.org/structure/1ABT}{1ABT}, Solution NMR structure) is a potent neurotoxin found in the venom of the Taiwanese many-banded krait snake, \textit{Bungarus multicinctus}. This small polypeptide (78 amino acids) acts as a highly specific and irreversible antagonist of the nicotinic acetylcholine receptors (nAChRs) at the neuromuscular junction, blocking acetylcholine binding and leading to muscle paralysis. The solved structure displayed in Figure \ref{fig:benchmark_structures} contains alpha-bungarotoxin (BGTX), and a synthetic dodecapeptide (alpha 185-196) corresponding to a functionally important region on the alpha-subunit of the nicotinic acetylcholine receptor (nAChR).
\end{itemize}

Another important point to note with regard to simulation benchmarks is the significant difference regarding how the JAX + JAX-MD workflow manages GPU utilization and memory compared to the Torch + ASE combination. Simulations on smaller systems will likely exhibit a larger advantage on the JAX versions as it maximizes GPU utilization, unlike the Torch + ASE versions. The relative difference should shrink with increasing system size as GPU capacity gets saturated. 

%Although \textit{mlip} does not yet provide extensive benchmarking tools, all the pre-trained models released have been validated to support stable MD simulations for systems of approximately 1,000 atoms (PDB 1ABT) for at least 1 nanosecond. See Table~\ref{tbl:models_overview} for more details on the inference speed and number of parameters.

\newpage
\begin{table}[h]
  \centering
    \caption{Speed performance on MD simulation of the different pre-trained models and backends. All tests were run on a single NVIDIA H100 GPU, and speed metrics are given in milliseconds per step, averaged over 1\,ns of simulation. 1UAO is a chignolin molecule with 138 atoms, while 1ABT is a system with 1205 atoms. All models included in the table achieved stable simulations on these benchmarks.}
    \vspace{0.2cm}
  \begin{tabular}{ll|crrr}
    \toprule
       Models             & Parameters & Systems & Jax + JAX-MD & Jax + ASE & Torch + ASE \\
    \midrule
    \multirow{2}{*}{MACE (large)} & \multirow{2}{*}{2,139,152} & 1UAO   & 6.3 ms/step & 11.6 ms/step & 44.2 ms/step \\
                          & &  1ABT & 66.8 ms/step & 99.5 ms/step & 157.2 ms/step \\
    \midrule
    \multirow{2}{*}{Modified MACE} & \multirow{2}{*}{2,203,504} & 1UAO   & 3.0 ms/step & 6.4 ms/step & n.a. \\
                          & &  1ABT & 26.6 ms/step & 48.7 ms/step & n.a. \\
    \midrule
    \multirow{2}{*}{ViSNet} & \multirow{2}{*}{1,137,922} & 1UAO & 2.9 ms/step & 6.2 ms/step & 33.8 ms/step \\
                          &  & 1ABT & 25.4 ms/step & 46.4 ms/step & 101.6 ms/step \\
    \midrule
    \multirow{2}{*}{NequIP} & \multirow{2}{*}{1,327,792} & 1UAO & 3.8 ms/step & 8.5 ms/step & 38.7 ms/step  \\
                          &  &  1ABT & 67.0 ms/step & 105.7 ms/step & 117.0 ms/step  \\
    \bottomrule
  \end{tabular}
  \label{tbl:models_overview}
\end{table}

\section{Roadmap for library development} \label{sec:roadmap}

We aim to release several updates to the current version (v0.1.3) in the coming months. These will include: 
\begin{itemize}
    \item Ability to incorporate total charge as input and models designed to predict charge-related labels (e.g. partial charges, dipole). \\
    \item Additional MLIP models with a priority to models that have received validation through additional research. Examples may include eSEN~\cite{fu2025learningsmoothexpressiveinteratomic} or GemNet~\cite{gasteiger2022gemnetocdevelopinggraphneural}. \\
    \item Accelerated backends for faster inference on steerable convolutions models (e.g. cuEquivariance, sparse kernel generators~\cite{bharadwaj2025efficientsparsekernelgenerator}). \\
    \item Additional functionalities (e.g. new loss functions, optimizers, layers for MACE, NequIP, or ViSNet). \\
    \item Release complementary libraries in which \textit{mlip} will be integrated for additional functionalities (e.g. coarse-grained methods for MLIP \cite{brunken2025universallyapplicabletunablegraphbased}, or BoostMD \cite{schaaf2024boostmdacceleratingmolecularsampling}). \\
    \item Training MLIP models on improved and more extensive datasets, for example, the Open Molecules 2025 (OMol25) dataset~\cite{levine2025openmolecules2025omol25}.
\end{itemize}

Our objective is for any addition to the library to remain open source.

% ATTENTION: I removed this next part for now because I find it a bit too tricky. We currently have
% absolutely no idea / strategy of how to handle contributions to the repo from the community - and this will likely remain tricky for a bit, even though I agree we should find a way eventually

% , and as such we encourage contributions from the community. Please refer to the relevant tutorial for methods to incorporate new elements into the library.

% Optional Acknowlegement
\section{Acknowledgements}

We would like to acknowledge beta testers for this library: Isabel Wilkinson, Nick Venanzi, Hassan Sirelkhatim, Leon Wehrhan, Sebastien Boyer, Massimo Bortone, Scott Cameron, Louis Robinson, Tom Barrett, and Alex Laterre.

\appendix
\newpage
\section{Appendix A - Complete set of model and training hyperparameters} \label{sec:appendix_A}
A complete description of each parameter can be found in the \textit{mlip} \href{https://instadeepai.github.io/mlip/api_reference/models/index.html}{model documentation}. The models were trained using very similar training strategies. Training was performed over 220 epochs with scheduled weights: energy (40) and forces (1000), flipped at epoch 115. An exponential moving average (EMA) with decay rate 0.99 was applied. The AMSGrad variant of Adam optimizer was used. The exponential moving average of the weights is taken at every training step.  We use 4000 warmup steps followed by 360000 transition steps. Gradient clipping was performed with a norm of 500, and no gradient accumulation was applied. The ViSNet and NequIP model training was performed using a Huber loss, while the MACE model training was performed using the MSE loss. See Table~\ref{tbl:nequip_params} for the hyperparameters used for the NequIP model, Table~\ref{tbl:visnet_params} for ViSNet, and Table~\ref{tbl:mace_params} for MACE. All training was done on NVIDIA H100 GPUs. Training took approximately 245 hours for NequIP, 158 hours for ViSNet and 266 hours for MACE. 

\begin{table}[ht]
  \centering
    \caption{NequIP model hyperparameters.}
  \begin{tabularx}{0.7\textwidth}{lX}
    \toprule
    \textbf{Parameter} & \textbf{Value} \\
    \midrule
    \texttt{num\_layers} & \texttt{5} \\
    \texttt{node\_irreps} &\texttt{64x0e + 64x0o + 32x1e + 32x1o + 4x2e + 4x2o} \\
    \texttt{l\_max} & \texttt{2} \\
    \texttt{num\_bessel} & \texttt{8} \\
    \texttt{radial\_net\_nonlinearity} & \texttt{swish} \\
    \texttt{radial\_net\_n\_hidden} & \texttt{64} \\
    \texttt{radial\_net\_n\_layers} & \texttt{2} \\
    \texttt{radial\_envelope} & \texttt{polynomial\_envelope} \\
    \texttt{scalar\_mlp\_std} & \texttt{4} \\
    \texttt{graph\_cutoff\_angstrom} & 5  \\
    \texttt{max\_n\_node} & 32  \\
    \texttt{max\_n\_edge} & 288 \\
    \texttt{batch\_size} & 16 \\
    \texttt{learning\_rate} & 0.002 \\
    \bottomrule
  \end{tabularx}
  \label{tbl:nequip_params}
\end{table}
\begin{table}[ht]
  \caption{ViSNet model hyperparameters.}
  \centering
  \begin{tabularx}{0.7\textwidth}{lX}
    \toprule
    \textbf{Parameter} & \textbf{Value} \\
    \midrule
    \texttt{num\_layers} & \texttt{4} \\
    \texttt{num\_channels} & \texttt{128} \\
    \texttt{l\_max} & \texttt{2} \\
    \texttt{num\_heads} & \texttt{8} \\
    \texttt{num\_rbf} & \texttt{32} \\
    \texttt{trainable\_rbf} & \texttt{False} \\
    \texttt{activation} & \texttt{silu} \\
    \texttt{attn\_activation} & \texttt{silu} \\
    \texttt{vecnorm\_type} & \texttt{None} \\
    \texttt{graph\_cutoff\_angstrom} & 5  \\
    \texttt{max\_n\_node} & 32  \\
    \texttt{max\_n\_edge} & 288 \\
    \texttt{batch\_size} & 16 \\
    \texttt{learning\_rate} & 0.0001 \\
    \bottomrule
  \end{tabularx}
  \label{tbl:visnet_params}
\end{table}
\begin{table}[ht]
  \caption{MACE model hyperparameters.}
  \centering
  \begin{tabularx}{0.7\textwidth}{lX}
    \toprule
    \textbf{Parameter} & \textbf{Value} \\
    \midrule
    \texttt{num\_layers} & \texttt{2} \\
    \texttt{num\_channels} & \texttt{128} \\
    \texttt{l\_max} & \texttt{3} \\
    \texttt{node\_symmetry} & \texttt{3} \\
    \texttt{correlation} & \texttt{2} \\
    \texttt{readout\_irreps} & \texttt{["16x0e","0e"]} \\
    \texttt{num\_readout\_heads} & \texttt{1} \\
    \texttt{num\_bessel} & \texttt{8} \\
    \texttt{activation} & \texttt{silu} \\
    \texttt{radial\_envelope} & \texttt{polynomial\_envelope} \\
    \texttt{graph\_cutoff\_angstrom} & 5  \\
    \texttt{max\_n\_node} & 32  \\
    \texttt{max\_n\_edge} & 288 \\
    \texttt{batch\_size} & 64 \\
    \texttt{learning\_rate} & 0.01 \\
    \bottomrule
  \end{tabularx}
  \label{tbl:mace_params}
\end{table}

\newpage
\section{Appendix B - SPICE1 training of MACE medium} \label{sec:appendix_B}

As discussed in the main body of the paper, we also trained a version of the MACE architecture of the library on a curated version of SPICE1 \cite{moore_spice1} with the same parameters as the original MACE-OFF medium model \cite{kovács2025maceofftransferableshortrange}. Overall, we achieved validation performance equivalent to MACE-OFF on SPICE1. We also present in Table \ref{tbl:mace_medium_overview} the runtime metrics on 1UAO and 1ABT. Likewise, the implementation of the ASE wrapper around the original torch version of MACE-OFF is our own, and it should not be considered representative of the performance of the original authors' code. 

\begin{table}[ht]
  \centering
    \caption{Speed performance on MD simulation of the MACE medium pre-trained models and backends. All tests were run on a single NVIDIA H100 GPU and speed metrics are given as milliseconds per step averaged over 1\,ns of simulation. 1UAO is a chignolin molecule with 138 atoms, while 1ABT is a system with 1205 atoms. Both models achieved stable simulation, and are both trained on a curated version of SPICE1 \cite{moore_spice1}.}
  \begin{tabular}{ll|crrr}
    \toprule
       Models             & Parameters & Systems & Jax + JAX-MD & Jax + ASE & Torch + ASE \\
    \midrule
    \multirow{2}{*}{MACE (medium)} & \multirow{2}{*}{1,911,568} & 1UAO  & 3.6 ms/step & 7.9 ms/step & 30.7 ms/step \\
                          & &  1ABT & 31.0 ms/step & 64.6 ms/step & 104.9 ms/step \\
    \bottomrule
  \end{tabular}
  \label{tbl:mace_medium_overview}
\end{table}

\section{Appendix C - Modified MACE model}

In this Appendix, we introduce a modified version of MACE that exhibits similar performances to the original MACE model while significantly reducing the associated inference time. A key component of MACE is the so-called Symmetric Contraction (SC), a costly node-wise operation that computes the tensor product of node features with themselves $\nu$ times, where $\nu$ is referred to as the correlation order. The choice of $\nu$ depends on a speed-accuracy trade-off, with a correlation order $\nu=3$ leading to a slower but more accurate model than $\nu=2$. Here we propose two simple yet effective modifications of the original MACE model that overcome this tradeoff, allowing to reach an accuracy close the one of our vanilla MACE model with $\nu = 3$ at an inference speed similar to the one obtained for $\nu = 2$.

In the following we describe our modifications by referring to the equations of the original MACE paper and adopt the same notations. Our first modification consists in applying a gating to the node features of node $i$ using the scalar produced during the SC. More precisely, we first define the following gating weights
\begin{equation}
\alpha^{(t)}_{Z_i kL, \eta_{\nu}} = \sum_{\tilde{\eta}_{\nu}}W^{\tilde{\eta}_{\nu}}_{Z_i k\tilde{k}, \eta_\nu}\mathbf{B}^{(t)}_{i,\tilde{\eta}_{\nu}\tilde{k}00} + b_{Z_ik,\eta_\nu}
\end{equation}
where $\mathbf{B}^{(t)}_{i,\tilde{\eta}_{\nu}\tilde{k}00}$
are the scalar features at the output of the SC and $W^{\tilde{\eta}_{\nu}}_{Z_i k\tilde{k}, \eta_\nu}, b_{Z_ik,\eta_\nu}$ are specie-wise learnable mixing weights and biases that depends on node $i$ through its atomic specie $Z_i$. Then, we gate the node features with the weights $\alpha^{(t)}_{Z_i kL, \eta_{\nu}}$, which amount to replace Eq. (11) defining the node messages with the following one:
\begin{equation}
m^{(t)}_{i,kLM} = \sum_{\nu}\sum_{\eta_{\nu}} \alpha^{(t)}_{Z_i kL, \eta_{\nu}}W^{(t)}_{Z_i kL, \eta_{\nu}} \mathbf{B}^{(t)}_{i,\eta_{\nu}kLM}\,.
\label{eq:nodes_gating}
\end{equation}
We found that this modification accounted for most of the improvement in our model. Importantly, this modification increases the body-order of the node-features while avoiding an increase in the correlation order, at a computational cost negligible compared to that of SC.

The second modification relates to the interaction of a node with its neighbors as encapsulated in Eq. (8) of the original MACE paper. We generalize this interaction term as to let it depend explicitly on the species of both the sender and receiver nodes. To do so, we first embed the specie of each node $i$ into a vector of learnable scalar features
\begin{equation}
s_{i} := (a^{1}_{Z_i}, \dots, a^{d}_{Z_i})^{T} \in \mathbb{R}^{d}\,,
\end{equation}
with $d=8$. Then, we build scalar feature vectors on each edge $(i,j)$ as
\begin{equation}
f_{ij} := \left[ s_{i}\, ||\, s_{j} \,||\, s_{i}\circ s_{j}\right] \in \mathbb{R}^{3d}\,,
\end{equation}
where $||$ and $\circ$ respectively denote concatenation and Hadamard product. The features $f_{ij}$ are then processed through a MLP to produce weights
\begin{equation}
\beta^{(t)}_{ij, k l_1 l_2 l_3} := \mathrm{MLP}(f_{ij})_{k l_1 l_2 l_3}\,.
\end{equation}
At last, Eq. (8) of the original MACE paper is to be replaced by the following equation:
\begin{equation}
A^{(t)}_{i, kl_3 m_3} = \sum_{l_1 m_1, l_2 m_2} C^{l_3 m_3}_{l_1 m_1, l_2 m_2} \sum_{j\in \mathcal{N}(i)} \beta^{(t)}_{ij, k l_1 l_2 l_3}R^{(t)}_{k l_1 l_2 l_3}(r_{ji})Y^{m_1}_{l_1}(\mathbf{\hat{r}}_{ji})\sum_{\tilde{k}}W^{(t)}_{k\tilde{k}l_2} h^{(t)}_{j,\tilde{k}l_2 m_2}\,.
\label{eq:se_message_passing}
\end{equation}

\end{document}